\documentclass[12pt]{article}
\usepackage{pic03}
\usepackage{hyperref}
\usepackage{url}
\usepackage{graphicx}

\begin{document}

\title{\bf HEAVY ION COLLISIONS}
\author{
Raimond Snellings        \\
{\em NIKHEF, Amsterdam, The Netherlands} \\
E-mail: Raimond.Snellings@nikhef.nl}
\maketitle

%
%
\begin{figure}[h]
\begin{center}
%
%
%
\vspace{4.5cm}
\end{center}
\end{figure}

\baselineskip=14.5pt
\begin{abstract}
Lattice QCD predicts a phase transition between hadronic matter and a
system of deconfined quarks and gluons (the Quark Gluon Plasma) at
high energy densities.
Recent results from the Brookhaven Relativistic Heavy Ion Collider
(RHIC) dedicated to the study of QCD at extreme densities will be
discussed and compared to measurements obtained at the CERN Super
Proton Synchrotron (SPS). 
\end{abstract}
\newpage

\baselineskip=17pt

\section{Introduction}
Quantum Chromo Dynamics (QCD) provides, as part of the standard model,
a very successful description of strong interaction processes
involving large momentum transfer. 
However, from first principles several important aspects
of QCD are still poorly understood. Examples are color confinement,
chiral symmetry restoration and the structure of the vacuum. 
Better understanding of these concepts can
be obtained if we are able to study quarks and gluons in a deconfined
state, the so-called Quark Gluon Plasma (QGP). 

Such a deconfined state might be created in the laboratory 
in heavy-ion collisions at the highest energies.
Theoretical guidance for this comes from Lattice QCD calculations.
Lattice QCD predicts that at an energy density $\epsilon \approx 1$ GeV/fm$^3$,
corresponding to a temperature of about 170 MeV, the system
undergoes a phase transition from nuclear matter to a deconfined
system of quarks and gluons. 

\begin{figure}[htb]
  \begin{center}
    \includegraphics[width=\textwidth]{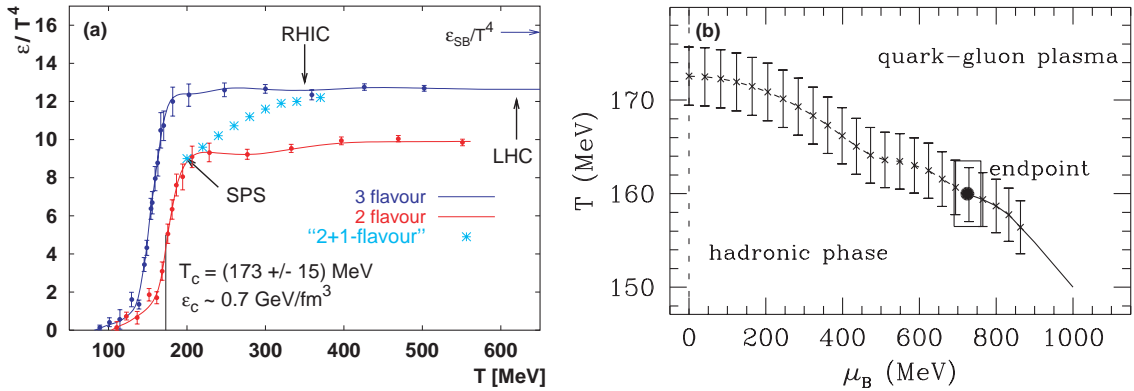}
    \caption{\it
      a) Energy density divided by T$^4$
      versus T at $\mu_B = 0$~\cite{Lattice}. b) Transition
      temperature as a function of $\mu_B$~\cite{LatticeFodor}. The
      dotted line illustrates the rapid crossover while the solid line
      illustrates the first order phase transition.
      \label{Lattice} }
  \end{center}
\end{figure}
Figure~\ref{Lattice}a shows the energy density divided
by the fourth power of the temperature, T, versus
the temperature from a lattice calculation~\cite{Lattice}. 
This figure shows that in between 150-200 MeV the energy density increases rapidly
which is indicative of a phase transition where at high temperature the
quarks and gluons become the relevant degrees of freedom. 
The figure also indicates were according to our current understanding the different
heavy-ion machines are located on this diagram. 
These calculations are done with zero baryon chemical potential,
$\mu_B$, reflecting the conditions of the early universe.
Small values of the chemical potential are obtained at RHIC collider
energies whereas at lower energies, e.g. AGS and SPS, the value of
$\mu_B$ is large.
In Fig.~\ref{Lattice}b the relation between the transition
temperature and the chemical potential from recent lattice
calculations~\cite{LatticeFodor} is shown. 
The calculation indicates that the transition temperature
decreases with increasing $\mu_B$ and furthermore that at low $\mu_B$
the transition from the hadronic phase to the QGP is a rapid crossover
(dotted line)
while at large $\mu_B$ a first order transition should take place
(full line).

\section{The Relativistic Heavy Ion Collider (RHIC)}

Heavy-ion physics entered a new era with the advent
of the Relativistic Heavy Ion Collider (RHIC) at
Brookhaven National Laboratory.
RHIC is a versatile collider providing collisions with different
ion species (ranging from protons to gold)
at a wide range of center of mass energies $\sqrt{s{_{_{NN}}}}$. 
In the three years of operation collisions were provided for Au+Au at 19.7, 130 and 200
GeV, p+p at 200 GeV and d+Au at 200 GeV. Note that the top center of mass energy
for p+p is 500 GeV at RHIC.
For details on the detectors (BRAHMS, PHENIX, PHOBOS, STAR),
see~\cite{RHICdetectors}.  

\section{Event Characterization}

Heavy ions are extended objects and the system created in a head-on
collision is different from that in a peripheral collision. Therefore,
collisions are categorized by their centrality. 
Theoretically the centrality is 
characterized by the impact parameter {\bf b} which, however, is not a
direct observable. 
Experimentally, the collision centrality can be inferred 
from the measured particle multiplicities if one assumes that this
multiplicity is a monotonic function of {\bf b}.
Another way to determine the event centrality is to measure the energy
carried by the spectator nucleons (which do not participate in the reaction) with
Zero Degree Calorimetry (ZDC). A large (small) signal in the ZDCs thus
indicates a peripheral (central) collision.

Two other measures of the centrality which are often used are
the number of wounded nucleons and the equivalent
number of binary collisions. 
These numbers are related to the impact parameter {\bf b} using a
realistic description of the nuclear geometry in a Glauber
calculation, see Fig.~\ref{nbin_nwounded}. 
Phenomenologically
it is found that soft particle production scales with the number of
participating nucleons whereas hard processes scale with the number of
binary collisions. 

\section{Low-$p_t$ Observables}

Examples of global observables which provide important information about the
created system are the particle multiplicity and the transverse energy. 
\begin{figure}[htb]
  \begin{minipage}[t]{0.48\textwidth}
    \begin{center}
      \includegraphics[width=\textwidth]{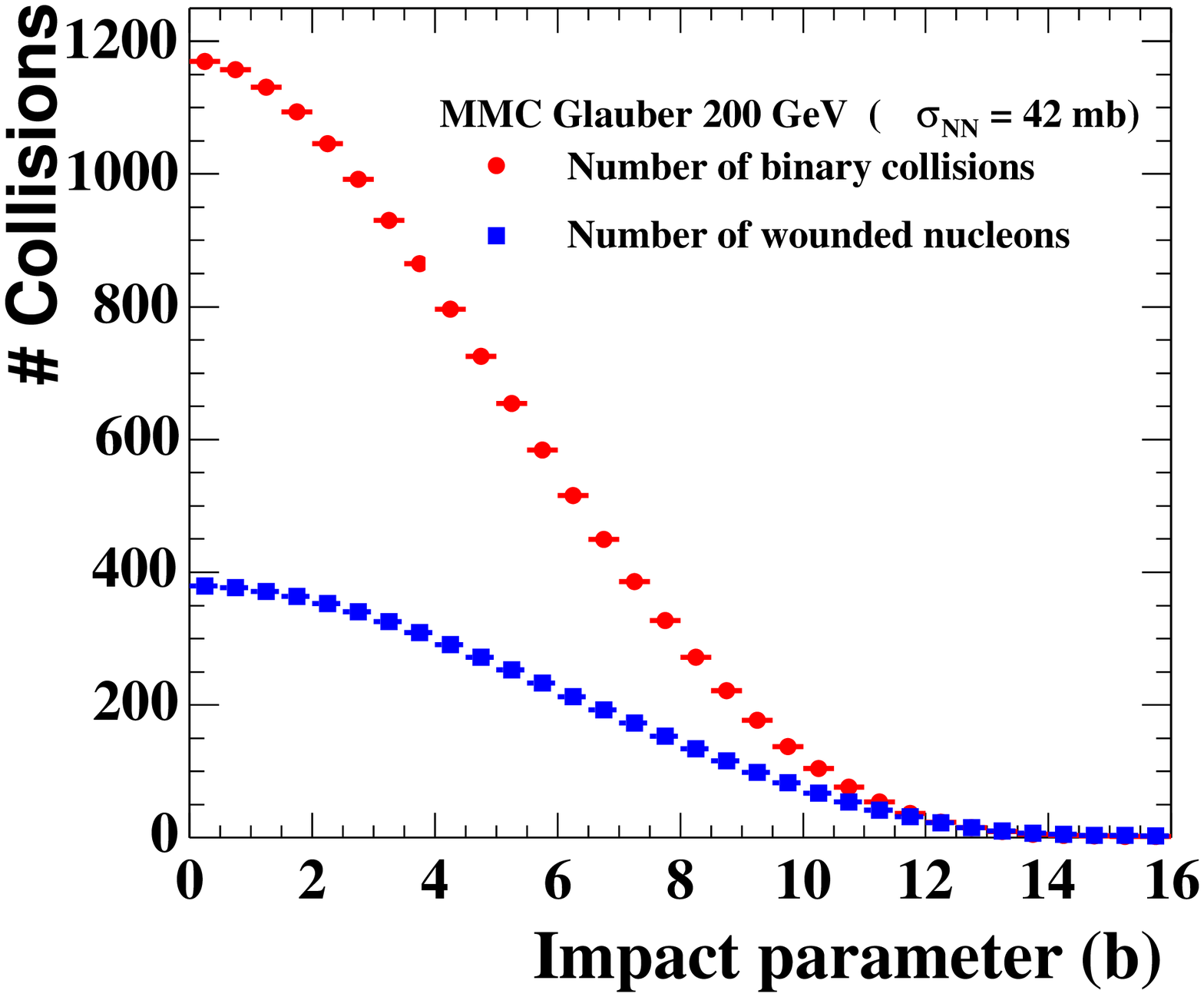}
      \caption{\it Number of wounded nucleons and binary collisions
        versus impact parameter.
      \label{nbin_nwounded} }
    \end{center}
  \end{minipage}
  \hspace{\fill}
  \begin{minipage}[t]{0.48\textwidth}
    \begin{center}
      \includegraphics[width=\textwidth]{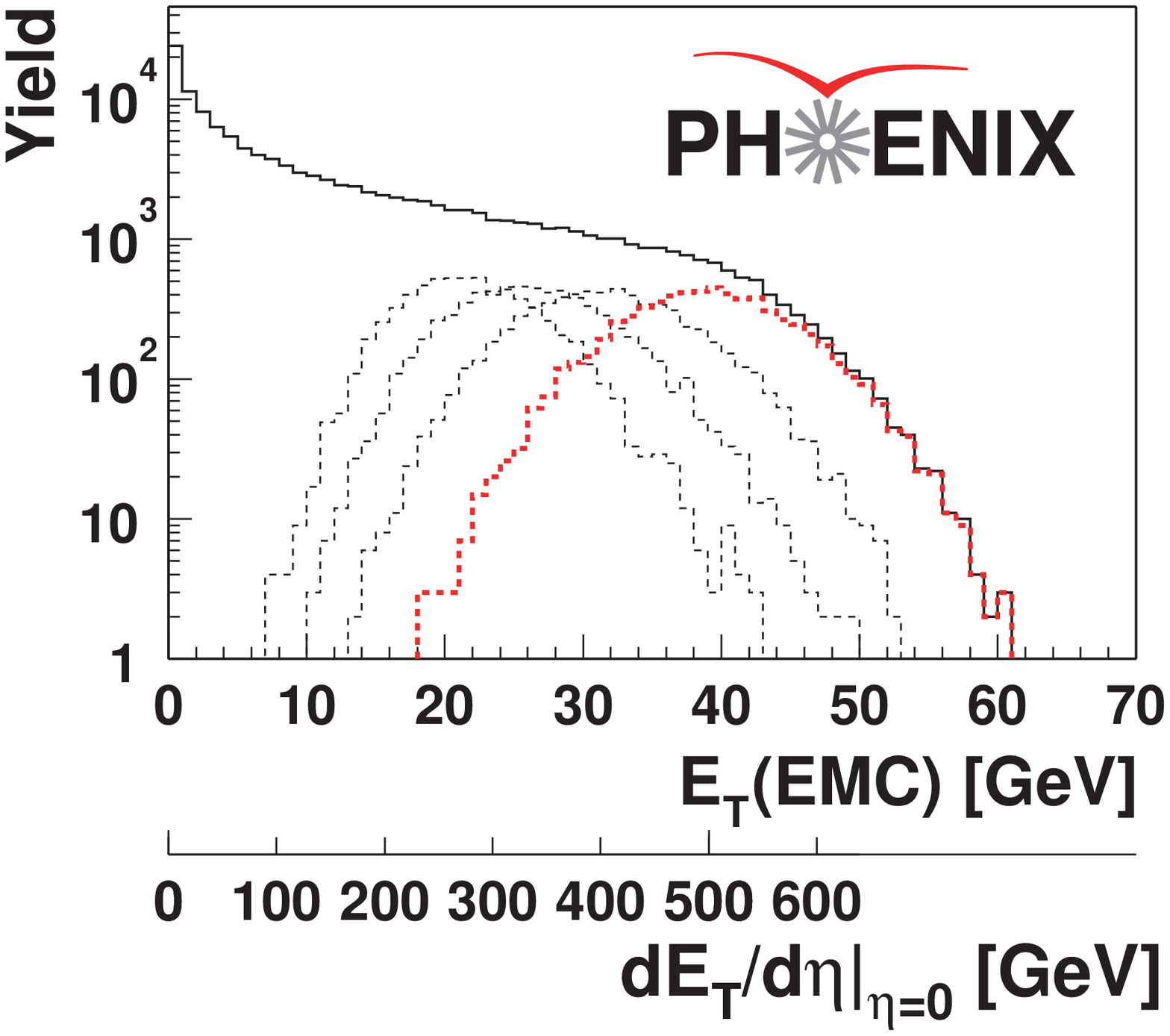}
      \caption{\it 
        Transverse energy as a function of centrality as measured by
        PHENIX~\cite{PHENIXet}.
      \label{phenixEt} }
    \end{center}
  \end{minipage}
\end{figure}
Figure~\ref{phenixEt} shows the transverse energy versus the collision
centrality as measured at
$\sqrt{s_{_{NN}}} = 130$~GeV by the PHENIX collaboration~\cite{PHENIXet}.
This
measurement allows for an estimate of the energy density as proposed by
Bjorken~\cite{bjorken} 
\[
\epsilon = \frac{1}{\pi R^2}\frac{1}{c \tau_0}\frac{dE_T}{dy},
\]
were $R$ is the nuclear radius and $\tau_0$ is the effective
thermalization time (0.2-1.0 fm/$c$). From the measured $\langle
dE_T/d\eta \rangle$ = 503 $\pm$ 2 GeV it follows that $\epsilon$ is 
about 5 GeV/fm$^3$ at RHIC. This is much larger than the critical
energy density of 1 GeV/fm$^3$ from Lattice QCD (see
Fig.~\ref{Lattice}). 

\begin{figure}[htb]
  \begin{center}
    \includegraphics[width=0.9\textwidth]{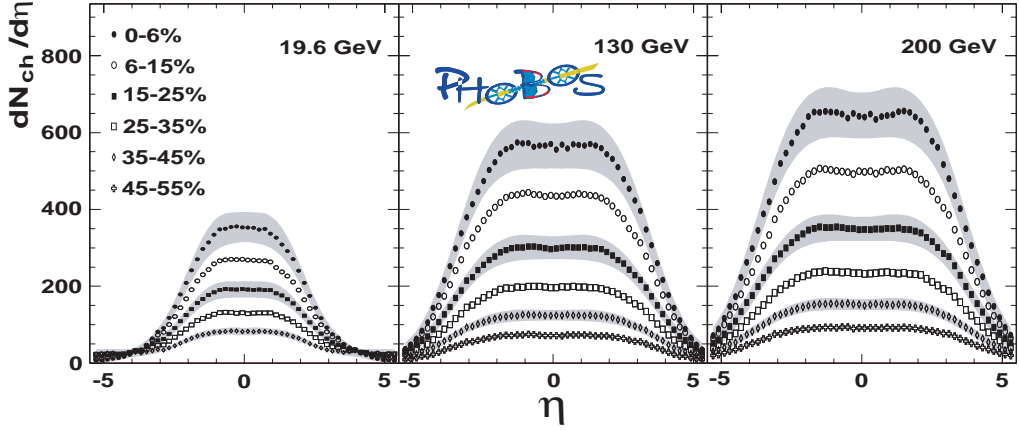}
    \caption{\it
      Multiplicity versus pseudo-rapidity for 19.6, 130 and 200
      GeV measured by PHOBOS~\cite{PHOBOSmultPRL}.
      \label{PHOBOSmult} }
  \end{center}
\end{figure}
Figure~\ref{PHOBOSmult} shows the charged particle multiplicity
distributions versus the pseudorapidity $\eta$ measured by PHOBOS at
three different energies~\cite{PHOBOSmultPRL}. 
The gross features of the particle multiplicity
distributions are described by a similar behavior of the tails
(limiting fragmentation) and a plateau at mid-rapidity 
consistent with a boost invariant region of $\Delta y \approx 1$. 
Notice that in total about 5000 charged particles are produced in the
most central Au+Au collisions at the top RHIC energy. 

\subsection{Particle Yields}

The integrated yield of the various particle species provides
information on the production mechanism and the subsequent
inelastic collisions. 
\begin{figure}[htb]
  \begin{center}
    \includegraphics[width=0.9\textwidth]{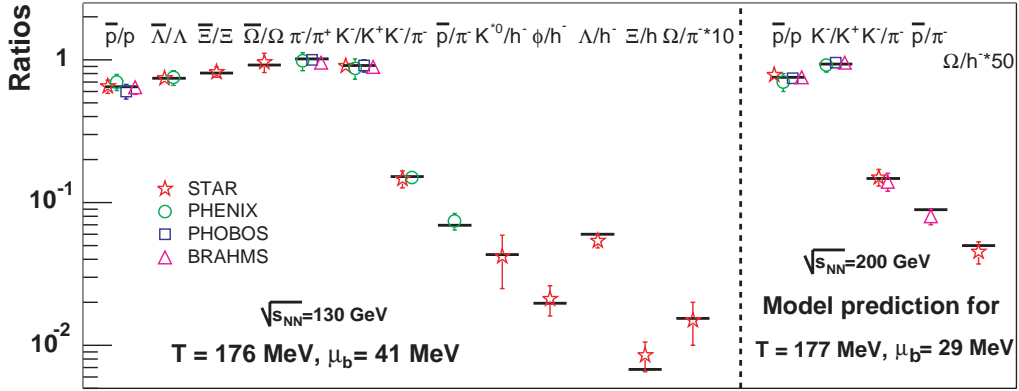}
    \caption{\it
      Particle yield ratios at RHIC compared with a thermal
      model~\cite{ParticleRatios}. 
      \label{thermalmodel} }
  \end{center}
\end{figure}
A very successful description of the relative particle yields is
given by the thermal model. In Fig.~\ref{thermalmodel} the particle
yield ratios measured at RHIC are plotted and compared to
values from a thermal model fit~\cite{ParticleRatios}. 
The results from the fit show
that all particles ratios are consistent with a single temperature and
single chemical potential in a thermal description. 
The temperature obtained in this way,
176 MeV, is called the chemical freeze-out temperature and the
measured value is very close to the phase boundary value in lattice
calculations (see Fig.~\ref{Lattice}b).  
\begin{figure}[htb]
  \begin{center}
    \includegraphics[width=\textwidth]{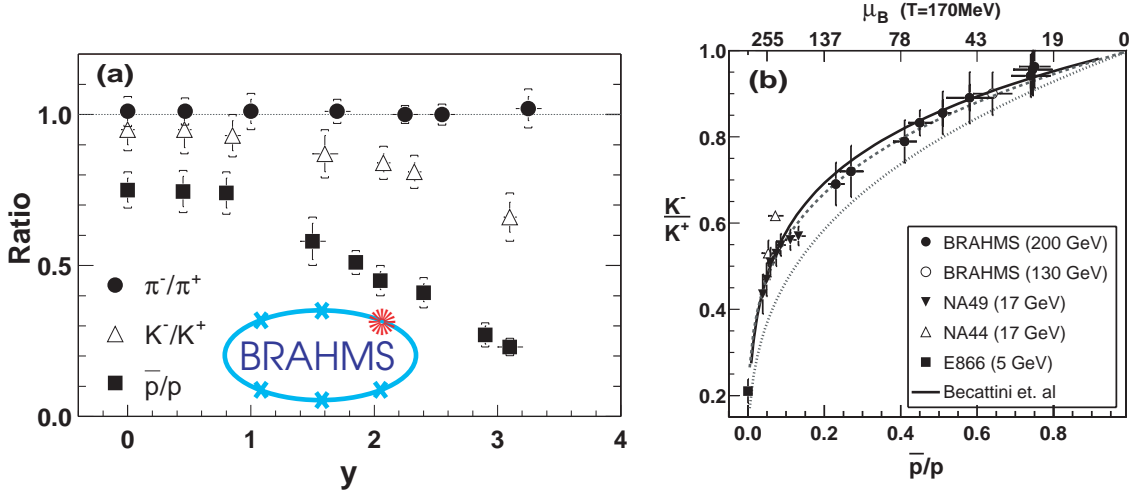}
    \caption{\it 
      a) Particle ratios versus rapidity measured by
      BRAHMS~\cite{BRAHMSratiosPRL}. 
      b) The ratio K$^-$/K$^+$ versus $\bar{\rm p}$/p or, equivalently, $\mu_B$.  
      \label{BrahmsRatios} }
  \end{center}
\end{figure}
Figure~\ref{BrahmsRatios}a shows the relative particle ratios of
pions, kaons and protons and their anti-particles versus
rapidity~\cite{BRAHMSratiosPRL}.  
For the heavier particles the ratio drops rapidly for $y > 1$.
Figure~\ref{BrahmsRatios}b shows the ratio of $K^-$/$K^+$ versus
$\bar{p}$/$p$ for AGS to RHIC energies. The decreasing ratio of
$\bar{p}$/$p$ as a function of rapidity can thus be understood 
from the changing baryon chemical potential at a constant chemical
freeze-out temperature.

\subsection{Spectra}
The particle spectra provide much more information than the
integrated particle yields alone. The particle yield as a function of
transverse momentum reveal the dynamics of the collision, characterized
by the temperature and transverse flow velocity of the system at
kinetic freeze-out.
Kinetic freeze-out corresponds to the final
stage of the collision when the system becomes so dilute that all
interactions between the particles cease to exist so that the 
momentum distributions do not change anymore. 
\begin{figure}[htb]
  \begin{center}
    \includegraphics[width=\textwidth]{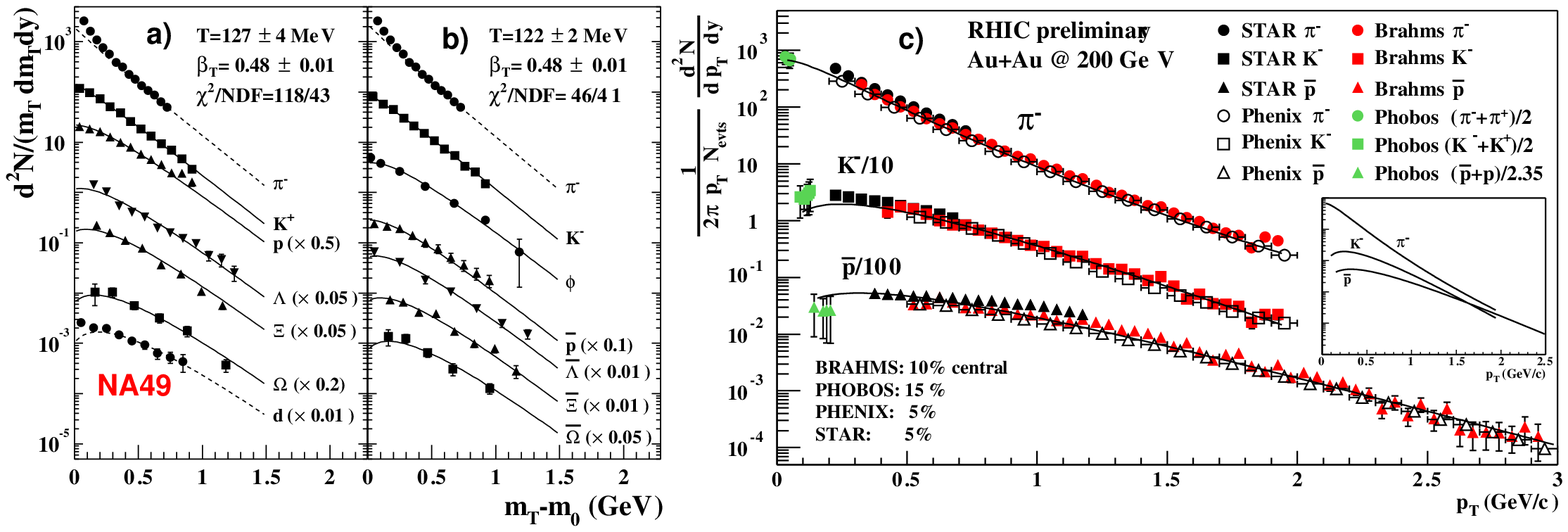}
    \caption{\it
      NA49 (SPS) and RHIC low-$p_t$
      spectra~\cite{Marco, Ullrich, BRAHMSspectra, PHOBOSspectra,
        STARspectra, PHENIXspectra}.  
      \label{lowptspectra} }
  \end{center}
\end{figure}
Figure~\ref{lowptspectra}a and b show the transverse momentum distributions
at $\sqrt{s}$ = 17~GeV from NA49~\cite{Marco}. 
The lines are a fit to the
particle spectra with a hydrodynamically inspired model (blast wave). 
The fit describes all the particle spectra rather well which shows
that these spectra can be characterized by the two
parameters of the model: a single
kinetic freeze-out temperature and a common transverse flow velocity.
Figure~\ref{lowptspectra}c shows the combined pion, kaon and proton
$p_t$-spectra from the four RHIC experiments. Also at these energies
it follows from a common fit to all the spectra that the system seems
to freeze-out with at a similar temperature and with a transverse flow
velocity as observed at SPS energies.

\subsection{Azimuthal Correlations with the Reaction Plane}

The nuclear overlap region 
in non-central collisions has an almond like shape with its longer axis
perpendicular to the reaction plane (the plane defined by the beam axis
and the impact parameter). This particular shape leads to a pressure
gradient which is different in and out of the reaction plane which, in
turn leads to azimuthally asymmetric
particle emission. 
The asymmetry can be described by:
\[
E\frac{{\rm d}^3N}{{\rm d}^3p} = \frac{1}{2\pi} 
\frac{{\rm d}^2N} {p_t {\rm d}p_t {\rm d}y} 
[ 1+ \sum_{n=1}^{\infty} 2 v_n {\rm cos}(n\phi)]
\]
where $\phi$ is the azimuthal angle with respect to the reaction
plane and the coefficient of the second harmonic, $v_2$, is called
elliptic flow. 
The magnitude of $v_2$ and its $p_t$ dependence allows
for the extraction of the kinetic freeze-out temperature and the
transverse flow velocity as function of emission angle. 
\begin{figure}[htb]
  \begin{minipage}[t]{0.48\textwidth}
    \begin{center}
      \includegraphics[width=\textwidth]{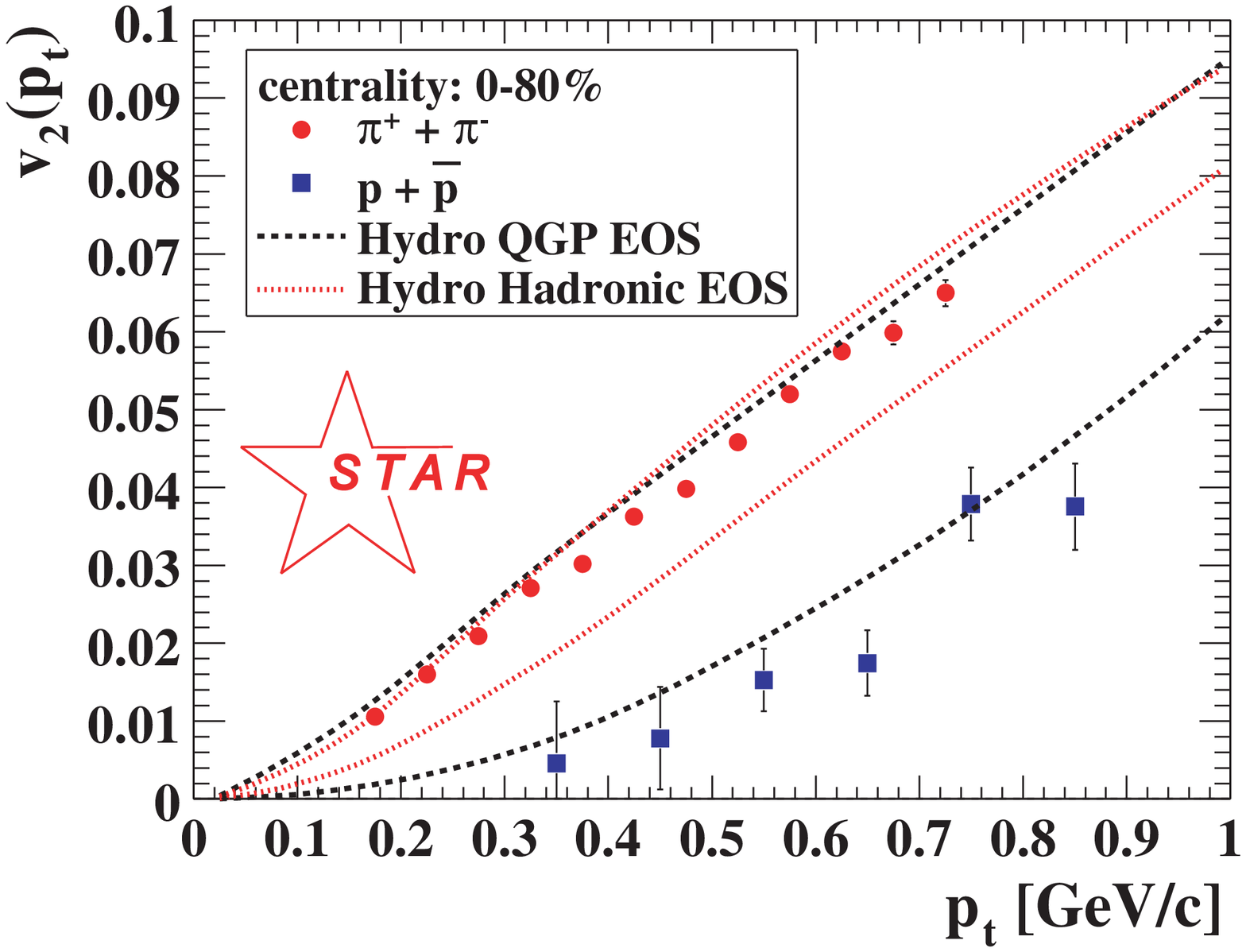}
      \caption{\it $v_2$($p_t$) for pions and protons at $\sqrt{s_{_{NN}}}
        =$ 130~\cite{STARpidflowPRL}. 
        The lines are hydrodynamical model calculations~\cite{HuovinenQM}.
      \label{Hydro_EOS} }
    \end{center}
  \end{minipage}
  \hspace{\fill}
  \begin{minipage}[t]{0.48\textwidth}
    \begin{center}
      \includegraphics[width=\textwidth]{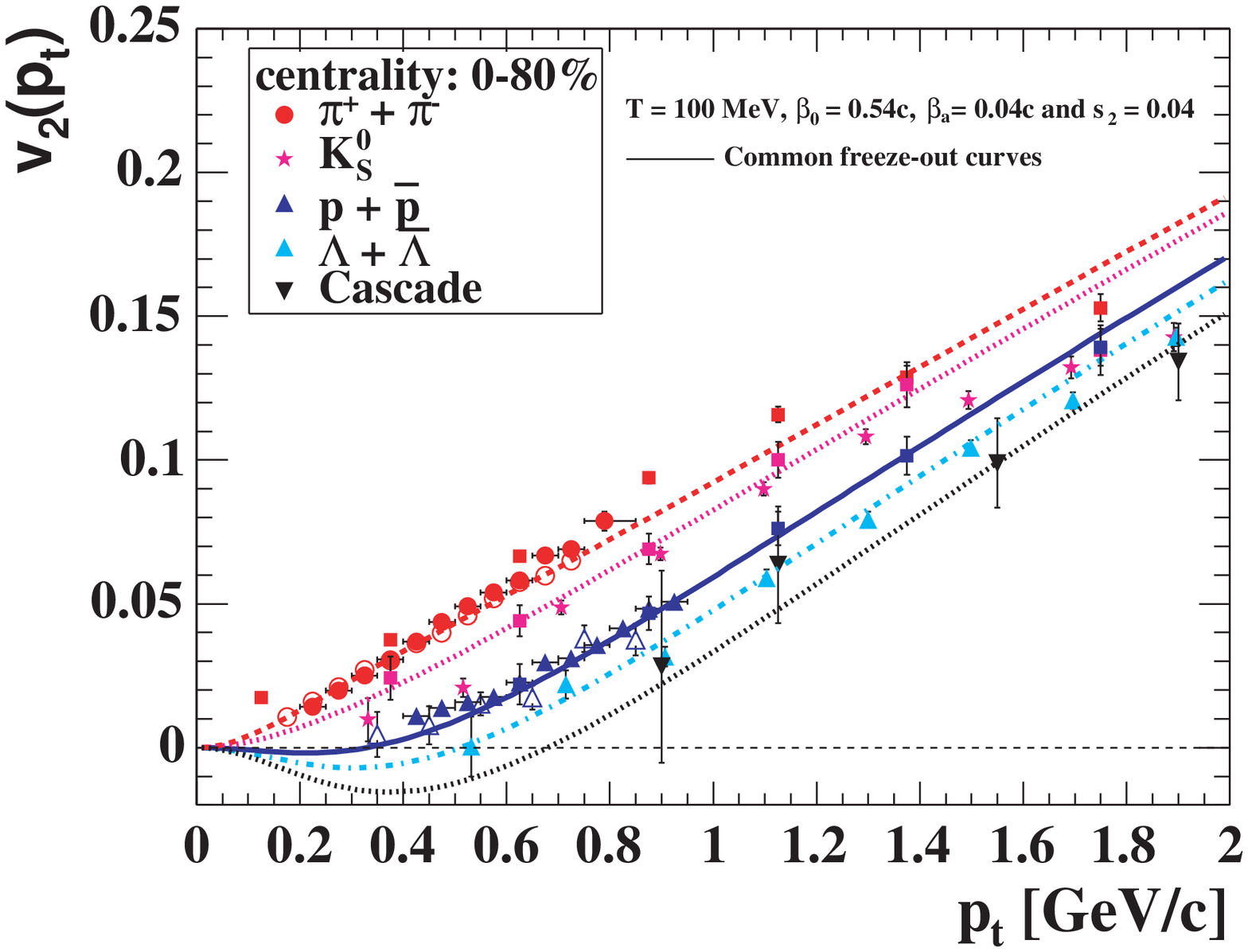}
      \caption{\it $v_2$($p_t$) for identified particles compared to a
        blast wave fit~\cite{STARpidflowPRL,
          STARklambdaflowPRL, SnellingsBreckenridge, PHENIXpidflowQM}. 
      \label{pid130_200} }
    \end{center}
  \end{minipage}
\end{figure}

Figure~\ref{Hydro_EOS} shows the measurement of $v_2$
versus $p_t$ for pions and protons plus antiprotons. 
Due to transverse flow the $p_t$ dependence of $v_2$ depends on the
particle mass as is evident from Fig.~\ref{Hydro_EOS}. Also
shown in this figure are hydrodynamical model calculations using
two different equations of state~\cite{HuovinenQM} corresponding a
hadron gas and a QGP. 
It is seen that the QGP EOS shows the best agreement with the data.
In Fig.~\ref{pid130_200} RHIC data on
$v_2$($p_t$)~\cite{STARpidflowPRL, STARklambdaflowPRL,
  SnellingsBreckenridge, PHENIXpidflowQM} for various particles are
compared to a hydrodynamical inspired blast wave fit.
The agreement of the data with this fit shows that the
$v_2$($p_t$) for all particles can be described in terms of a single
temperature and a $\phi$-dependent transverse flow
velocity.
Furthermore, the magnitude and $p_t$ dependence of the elliptic flow for the
various particles suggest strong {\it partonic} interactions in an early
stage of the collision and, perhaps, early thermalization of the system.

\section{High-$p_t$ observables}
\label{sec:highpt_observables} 

In heavy-ion collisions at RHIC, jets with transverse
energies above 40 GeV are produced in abundance, providing a detailed
probe of the created system. 
However the abundant soft particle production in heavy-ion collisions
tends to obscure the characteristic jet structures. 
At sufficient high-$p_t$ the
contribution from the tails of the soft particle production becomes
negligible and jets can be identified by their leading particles. 
It was proposed that a leading particle traversing a 
dense system would lose energy by induced gluon radiation (so called
jet-quenching~\cite{quenching}). The amount
of energy loss is in this picture directly related to the parton
density (mainly gluons at RHIC) of the created system. Currently there
are three observables sensitive to this energy loss as
discussed in the next two subsections.

\subsection{Single Inclusive Particle Yields}
As mentioned above, the single inclusive particle yield at sufficiently high-$p_t$ is
dominated by the leading particles from jets.
\begin{figure}[htb]
  \begin{center}
    \includegraphics[width=\textwidth]{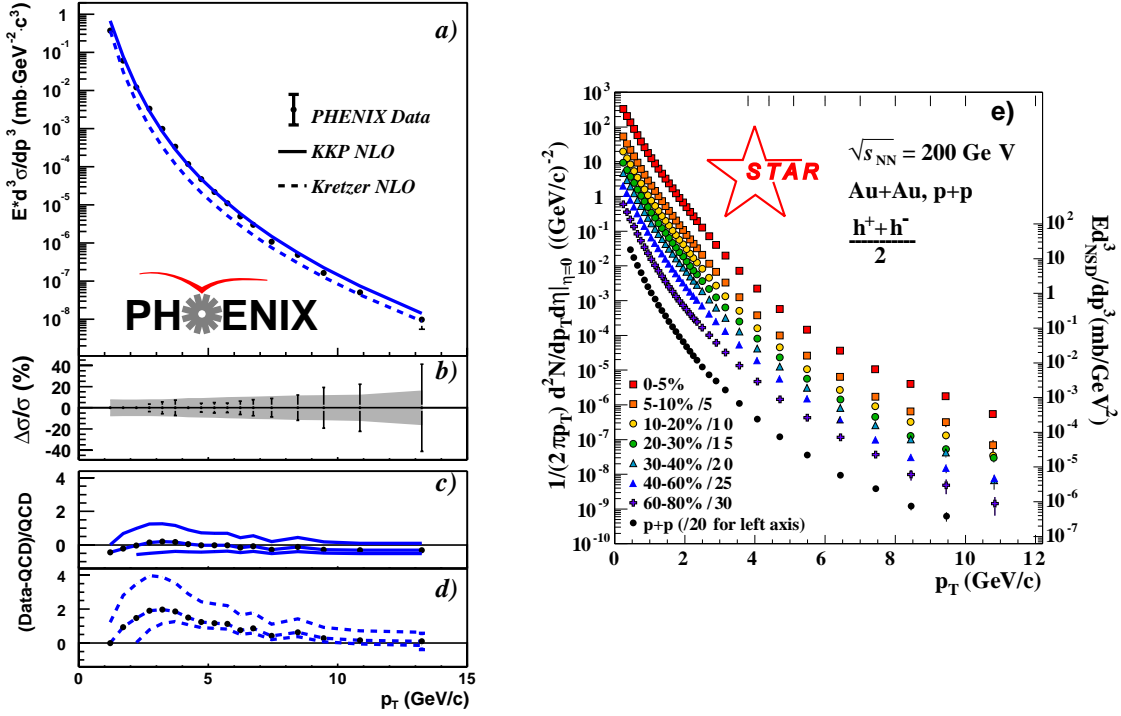}
    \caption{\it
      $\pi_{0}$ spectra in p+p~\cite{PHENIXpi0pp} and
      charged hadrons in p+p and Au+Au~\cite{STARraa}.
      \label{spectrahighpt} }
  \end{center}
\end{figure}
Figure~\ref{spectrahighpt}a shows the $\pi_0$ spectra as measured in
p+p at $\sqrt{s}$ = 200 GeV. In the same figure also two NLO QCD
calculations are shown. The ratio of the data to the theory shows
that in p+p the $\pi_0$ spectrum is well described. In
Fig.~\ref{spectrahighpt}e the charged hadron spectra measured in Au+Au
at $\sqrt{s_{_{NN}}}$ = 200 GeV and the p+p reference spectra at the same
energy are shown.
One of the observables suggested for measuring energy loss is the so
called nuclear modification factor defined by
\[
{R_{\rm AA}(p_t)} = 
\frac{{\rm d}^2\sigma_{\rm AA}/ {\rm d}y{\rm d}p_t}
{\langle N_{\rm binary} \rangle
   {\rm d}^2\sigma_{\rm pp}/ {\rm d}y{\rm d}p_t},
\]
where ${\rm d}^2\sigma_{\rm pp}/{\rm d}y{\rm d}p_t$ is the inclusive
cross section measured 
in p+p collisions and $\langle N_{\rm binary}\rangle$ accounts for the
geometrical scaling from p+p to nuclear collisions. In the case that a
Au+Au collision is an incoherent superposition of p+p collisions this
ratio $R_{\rm AA}$ would be unity. Energy loss and shadowing would
reduce this ratio below unity while anti-shadowing and the Cronin effect
would lead to a value above unity.
\begin{figure}[htb]
  \begin{center}
    \includegraphics[width=\textwidth]{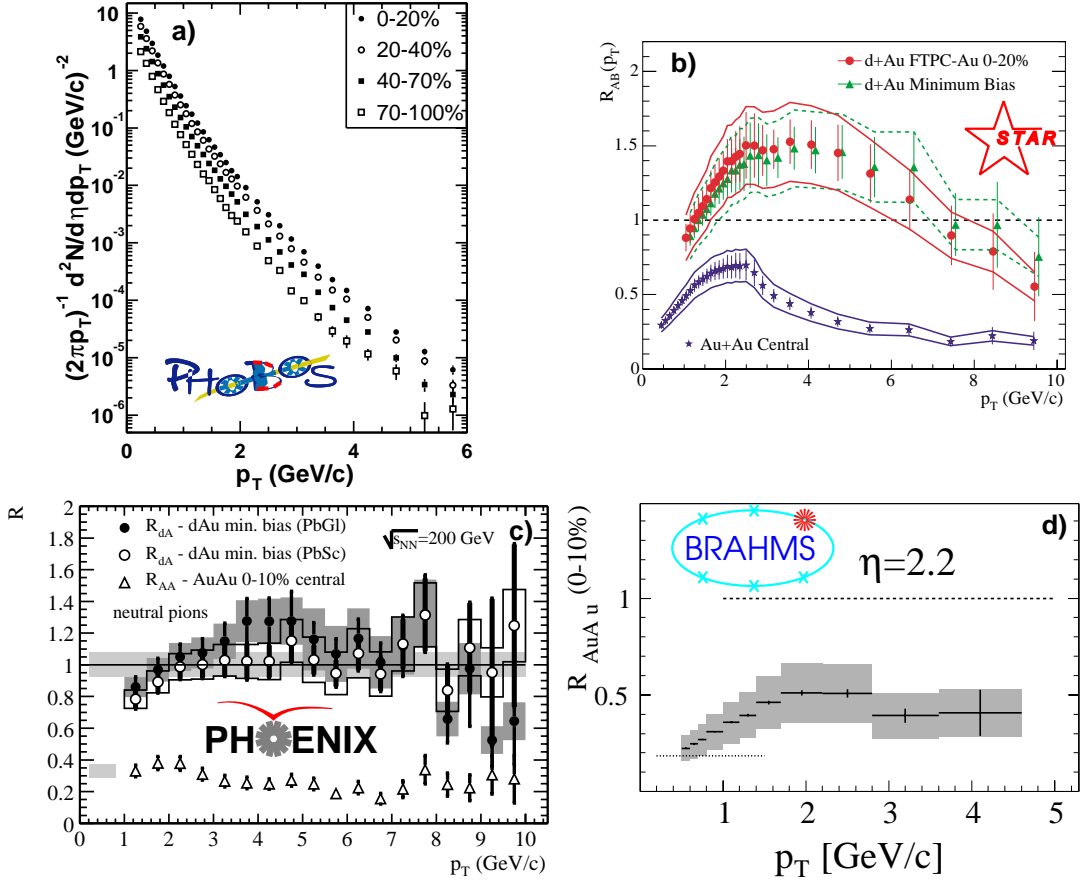}
    \caption{\it
      d+AU and Au+Au measurements from PHOBOS~\cite{PHOBOSraa},
      STAR~\cite{STARraa}, PHENIX~\cite{PHENIXraa} and BRAHMS~\cite{BRAHMSraa}.
      \label{RHICdAandAA} }
  \end{center}
\end{figure}
Figure~\ref{RHICdAandAA}b,c shows this ratio for charged particles and
$\pi_0$'s in central Au+Au
collisions at mid-rapidity. The ratio is well below one and at
high-$p_t$ the suppression is a factor of 5. At intermediate $p_t$ the
charged particles and $\pi_0$ are both suppressed; however the
magnitude differs by a factor of two. In Figure~\ref{RHICdAandAA}d
$R_{AA}$ is plotted at more forward rapidities showing that the
suppression also persists there. 

To discriminate between energy
loss and shadowing, d+Au collisions were measured. If the suppression is
due to shadowing it should also be observed in the d-Au
system. Figure~\ref{RHICdAandAA}a shows the d+Au spectra versus
centrality and Fig.~\ref{RHICdAandAA}b,c the nuclear
modification factor 
for charged particles and $\pi_0$, respectively. It is clear that in
d+Au interactions no suppression is observed. In fact, to the contrary,
a small enhancement is seen consistent with the Cronin effect. 
From this observation it follows that the observed suppression in
Au+Au collisions is due to final state interactions.
The magnitude of the observed suppression at the top RHIC energy
indicates, in the jet quenching picture, densities which are a factor 30 higher
than in nuclear matter. 

\subsection{Azimuthal Correlations}
In heavy-ion
collisions, azimuthal correlations between particles can be used to
study the effect of jet quenching in greater detail. 
The azimuthal correlations of two high-$p_t$
particles from jets are expected to show a narrow near-side
correlation and a broader away-side correlation.
However, in the case of strong jet quenching the 
away-side jet would suffer significant energy loss and would be suppressed.
\begin{figure}[htb]
  \begin{center}
    \includegraphics[width=0.95\textwidth]{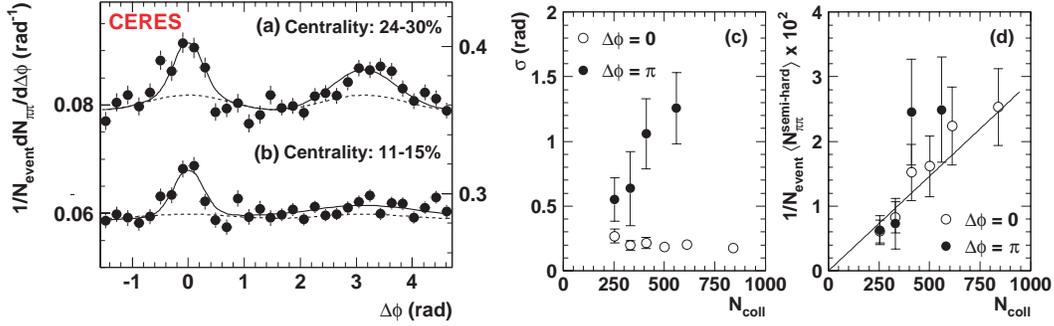}
    \caption{\it
      Azimuthal correlations measured at
      the SPS by CERES~\cite{CERESb2b}.
      \label{CERESbtob} }
  \end{center}
\end{figure}
Recently, CERES measured such a correlation function at the top SPS
energy. In Fig.~\ref{CERESbtob}a the nearside correlation (at
$\Delta\phi = 0$) shows a narrow peak consistent with the correlation
observed in jets. The away-side correlation peak is observed in more
peripheral collisions but disappears for more central collisions, see
Fig.~\ref{CERESbtob}b. 
Figure~\ref{CERESbtob}c shows that the width ($\sigma$) of
the near-side correlation peak stays constant as a function of
centrality, but that the away-side peak broadens for more central
collisions. The total integrated yield is the
same in the near and away-side peak (Fig.~\ref{CERESbtob}d). 
Therefore, the disappearance of the
away-side peak at the top SPS energy is interpreted as being due to
initial state broadening~\cite{CERESb2b}. 

\begin{figure}[htb]
  \begin{center}
    \includegraphics[width=0.95\textwidth]{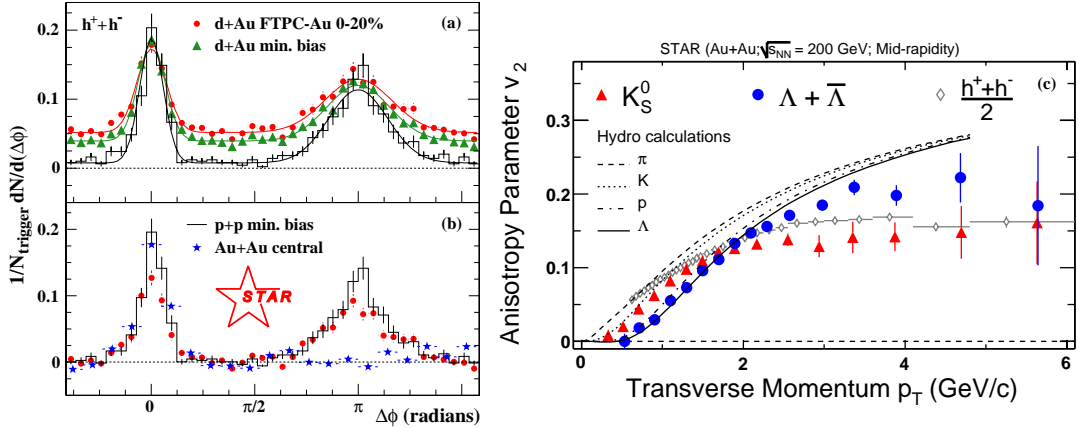}
    \caption{\it
      a,b) Back to back correlations~\cite{STARb2b} and c) elliptic flow
      parameter~\cite{STARklambdaflowPRL} at intermediate $p_t$.
      \label{STARcorr} }
  \end{center}
\end{figure}

The azimuthal correlations of high-$p_t$
particles (trigger particle 4 $<$ $p_t < 6$~GeV/$c$, associated particle
2~GeV/$c$ $<$ $p_t$ $<$ $p_t^{trig}$) measured in p+p collisions at
RHIC are shown as the histogram in Figure~\ref{STARcorr}b. 
The near-side and away-side peaks are clearly visable. 
The correlation function observed in central
Au+Au collisions (stars in Fig.~\ref{STARcorr}b) shows a similar near
side peak while the away-side peak has disappeared. 

To investigate if this is due to
initial state effects, the same analysis was done for d+Au
collisions. 
In Fig.~\ref{STARcorr}a the near and away-side peaks are
shown for minimum bias and central d+Au collisions compared to p+p.
The away-side correlation in d+Au is clearly observed even for the
most central collisions. Comparing the away-side correlation in p+p,
d+Au and Au+Au, Fig.~\ref{STARcorr}b, shows that the suppression only
occurs in Au+Au collisions and therfore is a final state effect as
expected from jet quenching.

The energy loss depends on the distance 
traversed through the dense medium by the partons. In a non-central
collision the distance will depend on the azimuthal angle with respect
to the reaction plane (see low-$p_t$ section). 
Because the hard scattering producing the di-jet has no correlation
with the reaction plane, 
an observed asymmetry in the high-$p_t$ particle emission will be due
to final state interactions (such as the jet quenching mechanism).
In Fig.~\ref{STARcorr}b the observed elliptic flow signal as a
function of $p_t$ is shown for charged particles, kaons and lamdas. It
is clear from this figure that the observed asymmetry is very large up
to the highest $p_t$ measured. 
Like the nuclear suppression factor $R_{AA}$, the elliptic
flow at intermediate $p_t$ depends on the particle species. This
could be due to an interplay between the soft hydrodynamical
behavior and the jet quenching, which would cause a mass dependence. 
However, more recently this
has been interpreted as a possible sign of particle production at
intermediate $p_t$ by parton coalescence. In that case it is not the
mass of the particle which is responsible for the splitting but rather
the number of constituent quarks (two for mesons and three for
baryons).  
A definitive test will be the measurement of elliptic flow of the
$\phi$-meson because in the coalescence interpretation it should have
an elliptic flow similar to the pions while in the hydrodynamical
interpretation is would have an elliptic flow value similar to the
proton. 
  
\section{Conclusions}
The first three years of RHIC operation have provided a wealth of
interesting data. We have seen that:
\begin{itemize}
\item
Particle yields indicate a chemical freeze-out of the system near the
phase boundary;
\item
Identified particle spectra are consistent with boosted thermal
distributions and
identified particle elliptic flow shows remarkable agreement with ideal
hydrodynamical calculations based on a QGP equation of state;
\item
The particle yield at high-$p_t$ is suppressed compared to
proton-proton reference data. 
The fact that this
suppression does not occur in d+Au collisions shows that it is a final state effect,  
consistent with parton energy loss
in dense matter (jet quenching);
\item
The suppression at intermediate $p_t$ shows a particle dependence which
could be explained by particle production, at intermediate $p_t$,
by parton coalescence;
\item
The elliptic flow at intermediate $p_t$ is large and also shows a
particle dependence. Like above, this is consistent with energy loss
in dense matter and particle production via parton coalescence;
\item
In the most central events the high-$p_t$ back to back correlations are
consistent with zero. Such disappearance of the away-side jets is 
expected in the case of very strong energy loss in a dense medium.
\end{itemize}

All these observations, taken together, are consistent with the 
creation of a very dense and strongly interacting system in heavy-ion
collisions at RHIC energies. 
While all these
observations are consistent with the creation of a QGP, more
detailed knowledge of QCD at high densities and
temperatures is required. This poses a formidable challenge for theory
but will be crucial for the detailed interpretation of the present and
future data taken at RHIC and LHC.

\end{document}